\documentclass[11pt]{article}
\usepackage{geometry}	
\usepackage{graphicx} 
\usepackage{setspace}\spacing{1.213}	
\usepackage[bottom]{footmisc}		
\usepackage{rotating,float}		
\usepackage[justification=centering,font=small]{caption} 
\usepackage{amsmath,amssymb}\allowdisplaybreaks[1] 
\usepackage{appendix} 
\usepackage{placeins}
\usepackage[bookmarks=true]{hyperref} 
\usepackage{bookmark}
\usepackage{url}
\usepackage{caption}
\usepackage{subcaption}
\usepackage[small,compact]{titlesec}  

\usepackage{xcolor}	
\hypersetup{
	colorlinks,
	linkcolor={red!50!black},
	citecolor={blue!50!black}}

\usepackage[authoryear,round]{natbib} 

\usepackage{amsthm}
\theoremstyle{plain}

\theoremstyle{definition}

\newtheoremstyle{indenteddefinition}
	{}
	{}
	{\hangindent=2em}
	{}
	{\bfseries}
	{.}
	{.5em}
	{}
\theoremstyle{indenteddefinition}

\newcounter{assumptiongroup}\stepcounter{assumptiongroup}

\newcommand{\E}{{ \mathbb{E} }}
\newcommand{\Prob}{{ \mathbb{P} }}
\newcommand{\argmax}[1]{{ \underset{ #1 }{ \mathrm{arg}\, \mathrm{max} } \ }}
\newcommand{\action}{{ \mathrm{action} }}
\newcommand{\inaction}{{ \mathrm{inaction} }}
\newcommand{\thetab}{{ \boldsymbol{\theta} }}
\newcommand{\Xb}{{ \boldsymbol{X} }}
\newcommand{\Ib}{{ \boldsymbol{I} }}

\title{Starting Small: 
Prioritizing Safety Over Efficacy in Randomized Experiments
Using the Exact Finite Sample Likelihood\thanks{Comments very welcome. Previous versions of this paper have been circulated under different working paper numbers and different titles including ``Counting Defiers,''  ``A Model of a Randomized Experiment with an Application to the PROWESS Clinical Trial,'' and ``General Finite Sample Inference for Experiments with Examples from Health Care'' \citep{kowalski2019a,kowalski2019b}.  We extend
special thanks to Jann Spiess for extensive regular feedback and to Chuck Manski, Alex Tetenov, Toru Kitagawa, and Donald Rubin for encouraging us to use statistical decision theory and teaching us about it.  We also thank Don Andrews,  Josh Angrist, Susan Athey, Victoria Baranov, Steve Berry, Michael Boskin, Kate Bundorf, Xiaohong Chen, Victor Chernozhukov, Peng Ding, Brad Efron, Ivan Fernandez-Val, Michael Gechter, Matthew Gentzkow, Florian Gunsilius, Andreas Hagemann, Sukjin Han, Jerry Hausman, Han Hong, Guido Imbens, Daniel Kessler, Jonathan Kolstad, Ang Li, Aprajit Mahajan, Elena Pastorino, John Pepper,  Demian Pouzo, Edward Vytlacil, Stefan Wager, Christopher Walters, David Wilson, and seminar participants at the Advances with Fields Experiments Conference at the University of Chicago, the AEA meetings, the Essen Health Conference, Notre Dame, the Stanford Hoover Institution, UCLA, UVA, the University of Zurich, the Yale Cowles Summer Structural Microeconomics Conference, and the Y-RISE Evidence Aggregation and External Validity Conference for helpful comments.  Marian Ewell provided helpful practical information about randomization in clinical trials.  We thank Charles Antonelli,  Bennett Fauber, Corey Powell, and Advanced Research Computing at the University of Michigan, as well as Misha Guy, Andrew Sherman, and the Yale University Faculty of Arts and Sciences High Performance Computing Center.  Tory Do, Simon Essig Aberg, Bailey Flanigan, Pauline Mourot, Srajal Nayak, Sukanya Sravasti, and Matthew Tauzer provided excellent research assistance. 
}} 
\author{Neil Christy and Amanda Ellen Kowalski}
\date{July 18, 2024}

\begin{document}
	\maketitle
	\singlespacing
	\begin{center}
\end{center}

\begin{abstract}
\noindent  We use the exact finite sample likelihood and statistical decision theory to answer questions of ``why?'' and ``what should you have done?'' using data from randomized experiments and a utility function that prioritizes safety over efficacy. We propose a finite sample Bayesian decision rule and a finite sample maximum likelihood decision rule.  We show that in finite samples from 2 to 50, it is possible for these rules to achieve better performance according to established maximin and maximum regret criteria than a rule based on the Boole-Fr\'{e}chet-Hoeffding bounds.  We also propose a finite sample maximum likelihood criterion.  We apply our rules and criterion to an actual clinical trial that yielded a promising estimate of efficacy, and our results point to safety as a reason for why results were mixed in subsequent trials.    
\end{abstract}
\newpage

\section{Introduction} \label{sec:intro}

Suppose you have a new medical intervention.  You want to know if it works and if it is safe, so you run an experiment with two people.  You start small because you prioritize safety: it would be worse for you to administer an unsafe intervention than it would be to withhold an intervention that works.  The experiment has ended.  The person assigned intervention is alive, and so is the person assigned control.  Why? What should you have done---given the intervention to both or to neither?  Questions like these have attracted recent interest in the study of causal inference \citep{gelman2013, pearl2018, imbens2020}.

In the finite sample of two people, there are four possibilities for why you could have observed one person alive in intervention and another alive in control.  You recognize that one possibility is that both would have lived regardless.  However, you hold out hope for a second possibility - only the person in control would have lived regardless, and the intervention was efficacious for the person assigned intervention such that they were alive in intervention but would have died in control.  At the same time, you recognize that the intervention could have been unsafe for the person assigned control, such that they were alive in control but would have died in intervention.  If that were the case, there is a third possibility that the intervention was unsafe for the person in control and efficacious for the person in intervention.  There is also a fourth possibility that the intervention was unsafe for the person in control, and the person in intervention would have lived regardless.  Is there some way that you could use these possibilities to decide what you should have done? 

We propose that you could decide by combining the exact finite sample likelihood implied by the randomization in the experiment with  statistical decision theory in the style of \citet{manski2004}, \citet{manski2007}, \citet{stoye2012}, \citet{manski2018}, \citet{manski2019} and \citet{manski2021}.  Through this approach, we advance the literature on statistical decision theory in finite samples \citep{canner1970, manski2007admissible, schlag2007, stoye2007finite, stoye2009, tetenov2012} by applying an exact finite sample likelihood and considering asymmetric utility functions that prioritize safety over efficacy.  

We exploit sparsity and curvature in the exact finite sample likelihood function to propose a finite sample Bayesian decision rule and a finite sample maximum likelihood decision rule.  We evaluate the exact finite sample performance of these rules relative to other rules using established decision criteria. We demonstrate that in sample sizes from 2 to 50, for a utility function that is asymmetric in that it prioritizes safety over efficacy, our rules can often achieve strictly better performance according to the established maximin criterion than a recently developed rule based on the Boole-Fr\'{e}chet-Hoeffding bounds \citep{ben2024}. The intuition for the improved performance is that these bounds do not consider the data generating process from the randomization, so they are not sharp in finite samples.  We also propose a finite sample maximum likelihood criterion. Our finite sample maximum likelihood rule and criterion do not require a prior.  

We apply our proposed rules and criterion to an actual clinical trial of 28 people that examined the impact of high dose Vitamin C on patients with sepsis \citep{zabet2016}.  Results from the trial were very promising, showing a reduction in 28-day mortality that was statistically significant at the 2.6\% level in the finite sample.  However, results from 10 subsequent larger trials showed mixed results, yielding mixed results overall \citep{sato2021}.  Our analysis offers an additional approach to reconcile the mixed results across trials.  The maximizer of the finite sample likelihood in the trial, which is in some sense our ``best guess'' for why we observed what we did, indicates that while the intervention was efficacious for 21 people, it was unsafe for the remaining 7. The rule based on the Fr\'{e}chet bounds does not consider this possibility, providing intuition for why our finite sample maximum likelihood rule can achieve lower maximin performance than the rule based on the Fr\'{e}chet bounds in the sample of 28.       

In the next section, we introduce our model of a random experiment and derive the corresponding likelihood function.  In Section \ref{sec:decision}, we propose two novel decision rules that use the exact finite sample likelihood, and we propose an exact finite sample maximum likelihood criterion.  In Section \ref{sec:performance}, we compare the performance of our rules to established rules.  In Section \ref{sec:applicaiton}, we apply our rules to an actual clinical trial.

\section{The Exact Finite Sample Likelihood from a Randomized Experiment} \label{sec:model}

\subsection{Model and Notation}

Following the potential outcomes model of \citet{neyman1923}, \citet{rubin1974, rubin1977}, \citet{holland1986} and others, we ascribe each individual a binary potential outcome $y_I \in \{0, 1\}$ in intervention and $y_C \in \{0, 1\}$ in control.   Individuals are randomly assigned to intervention ($Z = I$) or control ($Z = C$), and one of their potential outcomes is revealed as the observed outcome $Y$:
\begin{align*}
Y = \boldsymbol{1}_{\{Z = I\}} (y_I) + \boldsymbol{1}_{\{Z = C\}} (y_C),
\end{align*}
where $\boldsymbol{1}_{\{ \cdot \}}$ is the indicator function.

An individual's realized outcome depends only on their own potential outcomes and their inclusion in the intervention or control arm, ruling out network-type effects through a ``no interference’’ \citep{cox1958} or ``stable unit treatment value’’ \citep{rubin1980} assumption.  Throughout, we define Y=1 as ``alive’’ and Y=0 as ``dead.’’

Under these assumptions, individuals fall into one of four ``principle strata’’ defined by their combination of potential outcomes \citep{frangakis2002}.  We refer to these four groups as those who would live regardless ($y_I=1$, $y_C=1$), those for whom the intervention would be efficacious ($y_I=1$, $y_C=0$), those for whom the intervention would be unsafe ($y_I=0$, $y_C=1$), and those who would die regardless ($y_I=0$, $Y_C=0$).  Let $\theta_{y_I, y_C}$ represent the total number of individuals in the experiment with potential outcomes $(y_I, y_C) \in \{0, 1\} ^ 2$.  The sum $\theta_{1, 1} + \theta_{1, 0} + \theta_{0, 1} +\theta_{0, 0} \equiv n$ is the sample size of the experiment.  We represent these four integers compactly as $\boldsymbol{\theta} \equiv \big( \theta_{1, 1}, \theta_{1, 0}, \theta_{0, 1}, \theta_{0, 0} \big)$.  The value $\boldsymbol{\theta}$ summarizes the joint distribution of potential outcomes within the sample.

Let $X_{I1}$ represent the number of individuals in the intervention arm observed with $Y=1$, $X_{I0}$ represent the number of individuals in the intervention arm with $Y=0$, $X_{C1}$ represent the number of individuals in the control arm with $Y=1$, and $X_{C0}$ represent the number of individuals in the control arm with $Y=0$.  These values constitute the data observed from the experiment, and we represent the data compactly with $\Xb = \big( X_{I1}, X_{I0}, X_{C1}, X_{C0} \big)$.

We assume that the number of individuals randomized into the intervention arm $m$ is fixed by the experimenter (often, $m = n/2$), and that each configuration of $m$ individuals in the intervention arm is equally probable, as though the experimenter draws $m$ names randomly out of a hat.  As we detail in  \ref{sec:app_likelihood}, the likelihood of any distribution of potential outcomes within the sample  given the experimental data $\Xb$ can be written as:
\begin{align}
\mathcal{L}( \thetab \mid \Xb ) = 
&= \sum_{i=1}^{\theta_{1,1}}
	\binom{ \theta_{1,1} }{ i } \nonumber\\
	&\qquad \qquad \times 
		\binom{ \theta_{1,0} }{ x_{I1} - i }  \nonumber\\
	&\qquad \qquad \times 
		\binom{ \theta_{0,1} }
			{ \theta_{1,1} + \theta_{0,1} - x_{C1} - i }  \nonumber\\
	&\qquad \qquad \times
		\binom{ \theta_{0,0} }
			{ m + x_{C1} + i - \theta_{1,1} 
				- \theta_{0,1} - x_{I1} }
	\bigg/ \binom{n}{m}  \label{eq:likelihood}
\end{align}

This likelihood function is equivalent to that appearing in \citet{copas1973}.  Note that this likelihood function varies with the joint distribution of the potential outcomes even when holding constant the marginal distributions of the potential outcomes.  That is, when both $\theta_{1, 1} + \theta_{1, 0}$ and $\theta_{1, 1} + \theta_{0, 1}$ are held constant, the likelihood function maintains some curvature.

Denote by $\hat{ \Theta }(\Xb)$ the set of distributions that maximize the likelihood given the experimental data $\Xb$:
\begin{align*}
\hat{ \Theta }(\Xb) = \argmax{ \boldsymbol{\theta} } \mathcal{L}( \boldsymbol{\theta} \mid \Xb )
\end{align*}

There are finitely many vectors of integers $\boldsymbol{\theta}$ that sum to the actual number of participants in the experiment $n$, so the set $\hat{ \Theta }(\Xb)$ is nonempty.  Denote the (sometimes unique) elements of this set as $\hat{ \boldsymbol{\theta} } \in \hat{ \Theta }( \Xb )$   (when the maximizer of the likelihood is unique, we sometimes write $\hat{ \boldsymbol{\theta} }( \Xb )$).  While the set of distributions that maximize the likelihood does not generically have a convenient analytical form, the integer programming problem can be solved in small samples by an exhaustive search.

\subsection{Discussion of the Exact Finite Sample Likelihood}

Our running example of an experiment with two people provides a minimal working example to demonstrate curvature in the likelihood.  Suppose one person is alive in intervention, and another is alive in control.  The maximizer of the likelihood function indicates that both people would have lived regardless. The intuition is simple.  It is most likely that the intervention was efficacious for both because even if the randomization had gone the other way such that the person assigned to intervention were assigned to control and vice versa, you would have seen the same thing---the person assigned intervention would be alive and the person assigned control would be dead.  The value of the likelihood is 1.  For all other possibilities, you would have seen something else if the randomization had gone differently, and you would have seen it 50\% of the time, so the value of the likelihood is 0.5.  Paraphrasing Chris Ferrie from the board book ``Statistical Physics for Babies,'' \citep{ferrie2017}, physicists refer to the number of ways that you could have seen what you have seen as entropy, and higher entropy explains why some things are more likely than others.  

In an experiment with two people, there are four other outcomes that we could have observed, and the unique maximizer of the likelihood function for each indicates that both people have the same type.  If the person in intervention is alive and the person in control is dead, it is most likely that the intervention would have been efficacious for both people.  If the person in control is alive and the person in intervention is dead, it is most likely that the intervention would have been unsafe for both people.  Finally, if the people in intervention and control are both dead, it is most likely that both would have died regardless.  

In larger samples, it is not always possible for all the people in the experiment to be of the same type.  But, as \citet{fisher1935} recognized, it is always possible for all the people in the experiment to be of two types---either the two types for which the intervention is efficacious and unsafe or the two types that would live or die regardless.  However, it need not be the case that the maximizer of the likelihood function includes only two types.  Indeed, ascribing each participant to one of two types sometimes implies that assignment to intervention or control within each type is highly imbalanced, while balance between intervention and control within a type is more likely: $N$ choose $M$ is maximized at $M=N/2$ (when $N$ is even). Maximization of the likelihood requires trading off between the higher likelihood of fewer types and the higher likelihood of balance within each type.   Just as people are more similar if they belong to fewer types,  people of the same type are more similar if they are assigned intervention and control at the same rate.

\section{Applying Decision Theory with Proposed Rules and Criterion} \label{sec:decision}

Our statistical decision theory approach requires three components: (1) a utility function for the decision maker that captures their preferences over giving the intervention to all or none of the participants; (2) a decision rule that maps the data the decision maker sees to the action they should have taken; and (3) a criterion for evaluating the performance of decision rules.

\subsection{Utilities that Prioritize Safety over Efficacy}

We suppose that the decision maker has a utility function $u$ defined over the joint distribution of potential outcomes within the sample  and a binary action $a \in \{ \action, \inaction \}$.  Note that the utility function is defined over the distribution of \emph{potential} outcomes rather than \emph{realized} outcomes, as in \citet{ben2024}.  We interpret action as giving the intervention to everyone, and we interpret inaction as giving the intervention to no one.  To fix ideas, throughout this paper, we focus on the following utility function:
\begin{align}
u( \boldsymbol{\theta}, \action ) = \frac{1}{n} \Big( \frac{1}{2} \theta_{1, 0} - \theta_{0, 1} \Big), \qquad
u( \boldsymbol{\theta}, \inaction ) = \frac{1}{n} \Big( \theta_{0, 1} - \frac{1}{2} \theta_{1, 0} \Big) \label{eq:util}
\end{align}
By dividing by the sample size $n$, we interpret these utility functions as utility-per-participant. We interpret the constant marginal utility with respect to the number of each type of individual as a social welfare function that is utilitarian---with respect to individuals' \emph{potential} outcomes.

By defining utility over potential outcomes rather than realized outcomes, we can encode a decision maker's preference for safety over efficacy.  Here, the decision maker prefers inaction in the presence of individuals for whom the intervention would be unsafe over action in the presence of individuals for whom the intervention would be efficacious: letting $\boldsymbol{\theta}_{ \mathrm{unsafe} } = (0, 0, 1, 0)$ be a sample distribution with a single participant for whom the intervention would be unsafe and $\boldsymbol{\theta}_{ \mathrm{efficacious} } = (0, 1, 0, 0)$ be a sample distribution with a single participant for whom the intervention would be efficacious, then
\begin{align*}
u\big( \boldsymbol{\theta}_{\mathrm{unsafe}},  \inaction \big) > u\big( \boldsymbol{\theta}_{\mathrm{efficacious}}, \action \big)
\end{align*}

This utility function also satisfies other sensible properties: conditional on taking action, utility is increasing in the number of individuals for whom the intervention would be efficacious and decreasing in the number of individuals for whom the intervention would be unsafe; and conversely, conditional on inaction, utility is increasing in the number of individuals for whom the intervention would be unsafe and decreasing in the number of individuals for whom the intervention would be efficacious.  Alternative utility functions defined over potential outcomes could also include utility from individuals who would live or die regardless, which we avoid here for simplicity.

\subsection{Decision Rules}

The decision maker does not know the underlying distribution of potential outcomes, but learns some information about the distribution through the observation of the experimental data $\Xb$, which can inform their decision of whether or not to take action.  We let $\mathcal{A}(\cdot)$ denote a mapping from observed experimental data into a probability of taking action; that is, for a binary random variable $A$ taking values in $\{ \action, \inaction \}$, $\mathcal{A}(\Xb) = \Prob\big( A = \action \mid \Xb \big)$.  We refer to these mappings as decision rules, and we note that our setup clearly allows for decision rules that prescribe stochastic actions.  

We propose two potential decision rules for the decision maker to follow: one based on the finite sample maximum likelihood estimate of the distribution of potential outcomes, and a second based on updating a Bayesian prior over the possible distributions of potential outcomes.  We also exposit two additional decision rules established in the literature for comparison: the first is an ``empirical success rule'' \citep{manski2004} based on the estimated average intervention effect, and the second is a rule based on maximizing utility within a subset of the possible distributions using the Boole-Fr\'{e}chet-Hoeffding copula bounds \citep{ben2024}.

The first rule we propose is the finite sample maximum likelihood rule.  When there is a unique maximizer of the likelihood, the rule takes the following form:
\begin{align*}
\mathcal{A}_{\mathrm{ML}}(\Xb) = 
\begin{cases}
	1,
		& u\big(\hat{\thetab}(\Xb), \action \big) > u\big(\hat{\thetab}(\Xb), \inaction \big), \\
	1/2,
		& u\big(\hat{\thetab}(\Xb), \action \big) = u\big(\hat{\thetab}(\Xb), \inaction \big),  \\
	0,
		& u\big(\hat{\thetab}(\Xb), \action \big) > u\big(\hat{\thetab}(\Xb), \inaction \big).
\end{cases} 
\end{align*}
This rule prescribes that the decision maker should choose their action as though the maximum likelihood estimate of the distribution of potential outcomes is indeed the true distribution.  More generally, when the MLE is not necessarily unique, then:
\begin{align*}
\mathcal{A}_{\mathrm{ML}} (\Xb) = 
	\frac{1}{ \big| \hat{\Theta}(\Xb) \big| }
	\sum_{ \hat{\thetab} \in \hat{\Theta}(\Xb) } \Bigg[
		\boldsymbol{1}_{ \big\{ u\big( \hat{\thetab}, \action \big) > u\big( \hat{\thetab}, \inaction \big) \big\} }
		+ \Big( \frac{1}{2}\Big ) \boldsymbol{1}_{ \big\{ u\big( \hat{\thetab}, \action \big) = u\big( \hat{\thetab}, \inaction \big) \big\} } \Bigg]
\end{align*}
This rule can be interpreted as establishing a uniform belief over the set of maximum likelihood estimates and choosing the probability of action to maximize expected utility.

The second rule we propose is a finite sample Bayesian rule.  This rule prescribes that the decision maker should choose the action that maximizes their posterior subjective expected utility, given a uniform prior over all possible values of $\thetab$:
\begin{align*}
\mathcal{A}_{\mathrm{Bayes}} (\Xb) = 
\begin{cases}
	1,
		& \E \big[ u( \thetab, \action) \mid \Xb \big]
			> \E \big[ u(\thetab, \inaction) \mid \Xb \big], \\
	1/2,& \E \big[ u( \thetab, \action) \mid \Xb \big]
			= \E \big[ u(\thetab, \inaction) \mid \Xb \big], \\
	0,	& \E \big[ u( \thetab, \action) \mid \Xb \big]
			< \E \big[ u(\thetab, \inaction) \mid \Xb \big],
\end{cases}
\end{align*}
where the expectation is taken with respect to the distribution of $\thetab \mid \Xb$, calculated by combining the uniform prior with the finite sample likelihood.  In principle, a Bayesian rule can be constructed given any prior over the possible distributions of potential outcomes.  We choose a uniform prior as a natural example.

In addition to our proposed rules, we also consider two other rules for comparison.  First, we consider the ``empirical success rule'' of \citep{manski2004}. In our setting, this rule simply instructs the decision maker to choose action whenever the estimated average intervention effect is positive, to choose inaction whenever the estimated average intervention effect is negative, and to choose action with probability 1/2 if the estimated average intervention effect is zero:
\begin{align*}
\mathcal{A}_{ES} (\Xb) = 
\begin{cases}
	1,
		& \frac{ X_{I1} }{ X_{I1} + X_{I0} } 
			> \frac{ X_{C1} }{ X_{C1} + X_{C0} }, \\
	1/2,
		& \frac{ X_{I1} }{ X_{I1} + X_{I0} } 
			= \frac{ X_{C1} }{ X_{C1} + X_{C0} }, \\
	0,
		& \frac{ X_{I1} }{ X_{I1} + X_{I0} } 
			< \frac{ X_{C1} }{ X_{C1} + X_{C0} }.
\end{cases}
\end{align*}

We also consider a ``Fr\'{e}chet rule,'' based on the Boole-Fr\'{e}chet-Hoeffding bounds \citep{boole1854, frechet1957, hoeffding1940, hoeffding1941}.  These so-called Fr\'{e}chet bounds restrict the set of possible joint distributions of random variables in terms of their marginal distributions, and have been applied frequently in the treatment effects literature (see, for example, \cite{balke1997}, \cite{heckman1997}, \cite{manski1997mixing}, \cite{tian2000}, \cite{zhang2003}, \cite{fan2010}, \cite{mullahy2018}, \cite{ding2019}, and \cite{tian2000}).  In the style of \citet{ben2024}, we construct a decision rule for utility functions defined over potential outcomes that utilize these bounds. 

To construct this Fr\'{e}chet rule, we first estimate the marginal distribution of the potential outcome in intervention 
$\Prob \big( Y_I = 1 \big) \equiv p_I$ 
using the share of individuals observed with outcome $Y=1$ in the intervention arm, 
$\hat{p}_I(\Xb)= \frac{ X_{I1} }{ X_{I1} + X_{I0} }$,
and we estimate the marginal distribution of the potential outcome in control 
$\Prob \big( Y_C=1 \big) \equiv p_C$
using the share of individuals observed with outcome $Y=1$ in the control arm,
$\hat{p}_C(\Xb)= \frac{ X_{C1} }{ X_{C1} + X_{C0} }$.
We then construct the set of joint distributions of potential outcomes consistent with these estimated marginal distributions according to the Fr\'{e}chet-Hoeffding copula bounds:
\begin{align}
\widehat{FH}(\Xb) =
	\bigg\{ p \in \mathbb{R}^4:\ 
		p_{0,1} \in&\ 
			\Big[ \max \big\{ - \big( \hat{p}_I(\Xb) - \hat{p}_C(\Xb), 0 \big) \big\},\ 
				\min \big\{ \big( \hat{p}_C(\Xb), 1 - \hat{p}_I(\Xb) \big) \big\}\Big], \nonumber\\
		p_{1,0} =&\ \hat{p}_I(\Xb) - \hat{p}_C(\Xb) + p_{0,1}, \nonumber\\
		p_{1,1} =&\ \hat{p}_C(\Xb) - p_{0,1}, \nonumber\\
		p_{0,0} =&\ 1 - \hat{p}_I(\Xb) - p_{0,1} \bigg\}  \label{eq:FH}
\end{align}
The Fr\'{e}chet rule then chooses the action that maximizes the minimum utility within this set:
\begin{align*}
\mathcal{A}_{\mathrm{F}}(\Xb) =
\begin{cases}
	1,
		& \min_{ p \in \widehat{FH}(\Xb) } u(p, \action)
			> \min_{ p \in \widehat{FH}(\Xb) } u(p, \inaction), \\
	1/2,
		& \min_{ p \in \widehat{FH}(\Xb) } u(p, \action)
			= \min_{ p \in \widehat{FH}(\Xb) } u(p, \inaction), \\		
	0
		& \min_{ p \in \widehat{FH}(\Xb) } u(p, \action)
			< \min_{ p \in \widehat{FH}(\Xb) } u(p, \inaction)
\end{cases}
\end{align*}
In \ref{sec:app_cutoff}, we show that, under our chosen utility function, the Fr\'{e}chet rule simplifies to a comparison of the estimated average intervention effect 
$\hat{p}_I(\Xb) - \hat{p}_C(\Xb)$
to a non-zero threshold (that varies with the estimated marginal distributions):
\begin{align*}
\mathcal{A}_{\mathrm{F}}(\Xb) =&\ 1 & 
\Leftrightarrow&\ & 
\hat{p}_I(\Xb) - \hat{p}_C(\Xb)
	>&\ \frac{1}{2} \min \big\{ \hat{p}_C(\Xb), 1-\hat{p}_I(\Xb) \big\}
\end{align*}

\subsection{Criteria for Comparing Decision Rules}

We consider a number of criteria for ranking decision rules that connect the decision maker's utility function, the decision maker's lack of knowledge of the true distribution of potential outcomes, and the data generating process of the experiment.  

To account for our potentially stochastic decision rules, we first introduce some additional notation: let 
$U\big( \thetab, \mathcal{A}(\Xb) \big)$
represent the decision maker's expected utility conditional on both the distribution of potential outcomes and the probability of choosing action given the observed data $\mathcal{A}(\Xb)$:
\begin{align*}
U\big( \thetab, \mathcal{A}(\Xb) \big) \equiv
\mathcal{A}(\Xb) u\big( \thetab, \action \big)
	+ ( 1 - \mathcal{A}(\Xb) ) u\big( \thetab, \inaction \big)
\end{align*}

First, we consider a decision maker who ranks decision rules by their worst-case expected utility:
\begin{align*}
V_{\mathrm{MM}}(\mathcal{A}) = 
\min_{\thetab} \E \big[ U\big( \thetab, \mathcal{A}(\Xb) \big)
	\mid \thetab \big],
\end{align*}
where the expectation is taken over all realizations of the experimental data $\Xb$. We refer to this well established preference ordering over decision rules as the maximin utility criterion.

We may alternatively assume that the decision maker ranks decision rules by another well established criterion: the decision maker first normalizes their expected utility from the rule by the best-case utility possible conditional on the joint distribution of potential outcomes (when using a loss function rather than a utility function, this quantity is known as ``regret'').  The decision maker then evaluates the rule based on the worst-case of this normalized expected utility:
\begin{align*}
V_{\mathrm{MMN}}(\mathcal{A}) = 
\min_{\thetab} \Big\{ 
	\E \big[ U\big( \thetab, \mathcal{A}(\Xb) \big)
		\mid \thetab \big]
	- \max_{a} u(\thetab, a) \Big\}
\end{align*}
We refer to this preference ordering as the maximin normalized utility criterion (which corresponds to the minimax regret criterion for utility instead of loss).

The maximin utility criterion has been criticized as too conservative \citep{savage1951}, while the maximin normalized utility (minimax regret) criterion is challenging to optimize and can result in undesirable ``no data'' decision rules, perhaps because of ``selective ignorance of likelihoods'' \citep{stoye2009}. In light of these criticisms, we also consider a Bayesian criterion and a novel maximum likelihood criterion.  For the Bayesian criterion, we assume that the decision maker carries a prior belief over the joint distribution of potential outcomes, encoded by the probability distribution $F(\thetab)$, and maximizes their total expected utility:
\begin{align*}
V_{\mathrm{B}}(\mathcal{A}) = 
\E_F \Big[
	\E \big[ U\big( \thetab, \mathcal{A}(\Xb) \big) \mid \thetab \big] \Big]
\end{align*}
Our proposed Bayesian rule can be shown to be optimal by this criterion when $F$ is the uniform prior.

Finally, we also consider a novel finite sample maximum likelihood criterion:
\begin{align*}
V_{\mathrm{ML}}(\mathcal{A}) = 
\min_{\boldsymbol{x}} \bigg\{
	\frac{1}{ \big| \hat{\Theta}(\boldsymbol{x}) \big| }
	\sum_{ \thetab \in \hat{\Theta}(\boldsymbol{x}) } \big[
		U\big( \hat{\thetab}, \mathcal{A}(\boldsymbol{x}) \big)
		- U\big( \hat{\thetab}, \mathcal{A}_{\mathrm{ML}}(\boldsymbol{x}) \big) 
	\big] \bigg\}
\end{align*}
For each possible realization of the experimental data $\boldsymbol{x}$, the maximum likelihood criterion compares the utility of taking action with probability 
$\mathcal{A}(\boldsymbol{x})$ 
to the utility of taking action with probability 
$\mathcal{A}_{\mathrm{ML}}(\boldsymbol{x}) ,$
given that the true distribution of potential outcomes is exactly the maximum likelihood estimate 
$\thetab(\boldsymbol{x}) .$  
When the likelihood is not uniquely maximized, the average value is taken across the maximizers. Decision rules are then ranked according to the worst-case realization $\boldsymbol{x}$.

The principle of maximum entropy as expounded by \citet{jaynes1957a, jaynes1957b}, who related statistical physics and the theory of information, provides motivation for our finite sample maximum likelihood criterion.  \citet{skyrms1987}, who discusses the principle of maximum entropy as a method for updating and supposing, credits \citet{kullback1951} with a similar idea and notes that the principle of maximum entropy has been applied in many fields.  Our understanding is that in some fields, the principle of maximum entropy has been used to derive a likelihood or an approximation of it.  In a randomized experiment, while we could derive expressions for entropy, we have the advantage of knowing the exact finite sample likelihood and using it directly.  An important advantage of the maximum likelihood decision rule and criterion over Bayesian decision rules and criteria is that they do not require specification of a prior.  In some sense, the maximum of the likelihood function is a prior chosen in a disciplined way, and the maximum likelihood criterion allows us to compare decision rules under that prior.  Our maximum likelihood criterion provides some contrast to the maximin and maximin regret criteria, which can sometimes compare rules based on effective priors that are very unlikely.

\section{Exact Finite Sample Performance of our Proposed Rules} \label{sec:performance}

\begin{figure}[!hbt]
	\caption{Performance of Decision Rules under the Maximin Utility Criterion}
	\centering
	\includegraphics[width=1\linewidth]{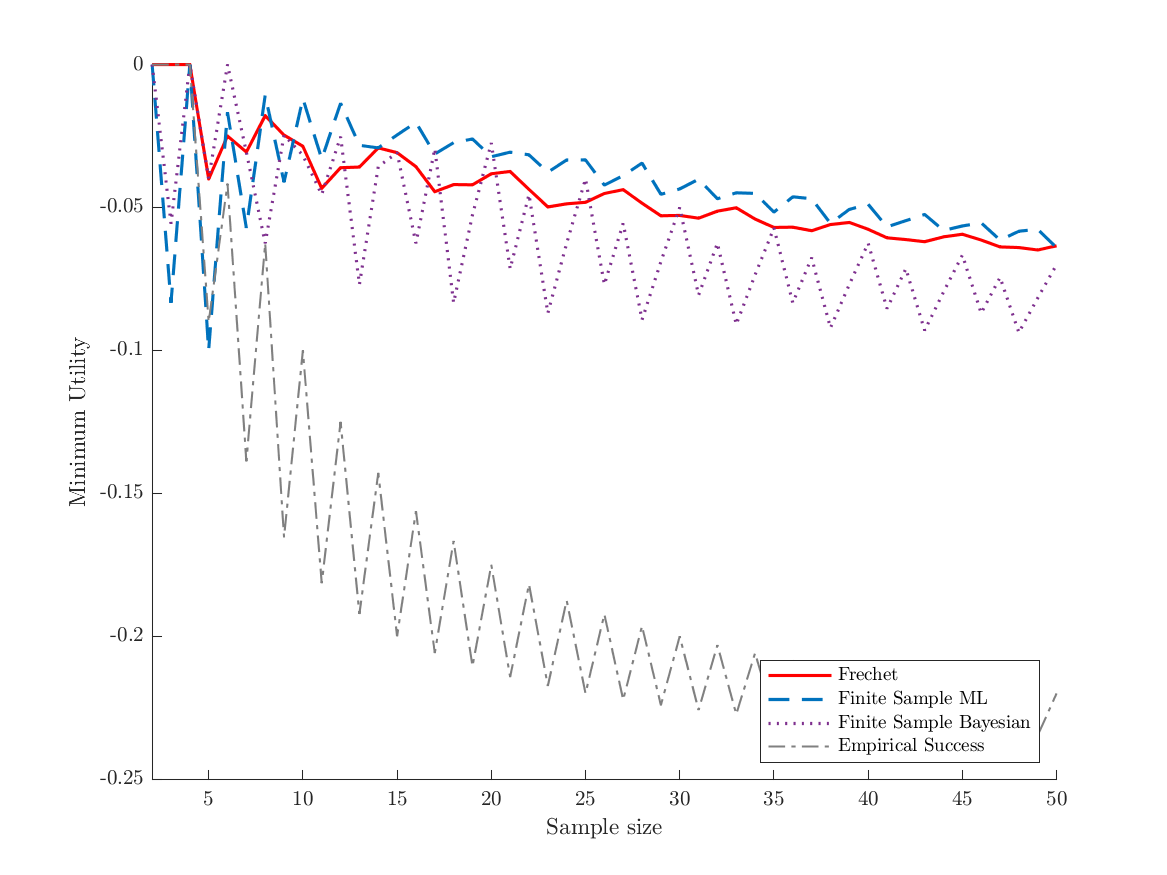}
	\label{fig:minU} 
\end{figure}

We evaluate the performance of our proposed rules under the criteria discussed in Section \ref{sec:decision} by explicitly calculating the expected utility achieved by each rule under each of the large (but finite) number of distributions of potential outcomes $\thetab$.  As shown in Figure \ref{fig:minU}, across all sample sizes from 2 to 50, our finite sample maximum likelihood rule often achieves higher minimum utility and therefore strictly outperforms the Fr\'{e}chet rule according to the established maximin criterion.  Our finite sample Bayesian rule with a uniform prior can also sometimes outperform the Fr\'{e}chet rule.  It is not surprising that the empirical success rule generally exhibits the worst performance because it is near optimal \citep{manski2004} only for symmetric utility functions that put equal weight on safety and efficacy in absolute terms.  We also note that, under our choice of utility function, a rule that prescribes randomizing between action and inaction regardless of what data is observed guarantees an expected utility of zero.  This rule would outperform any of the rules we consider in terms of the maximin criterion despite its ignorance of the data, which we consider an undesirable feature of this decision criterion.

Our interpretation of this exercise is that it shows that it is possible for our proposed finite sample maximum likelihood rule and our proposed finite sample Baysian rule to have better maximin performance than a rule based on the Fr\'{e}chet bounds in finite samples.  These performance measures are for a single utility function, and performance can and does vary with the choice of utility function.  However, it is perhaps counterintuitive that our rules could ever achieve better maximum performance than a rule based on bounds.

A sample of two people provides some intuition for why the Fr\'{e}chet rule can be beaten by our proposed rules that take the finite sample likelihood into account.  In our running example of two people in which one is alive in intervention and the other is alive in control, our rules consider all four possibilities that could explain why we observe what we observe, but the Fr\'{e}chet bounds only consider the possibility that both people would have lived regardless.  In an infinite population, if the intervention could have been unsafe for anyone, meaning that they would have been alive in control but dead in intervention, then the population would be large enough that someone would have been dead in intervention.  Therefore, if no one is dead in intervention in an infinite population, it is not possible that the intervention is unsafe for anyone.  Likewise, if no one is dead in control in an infinite population, then is it not possible that the intervention is efficacious for anyone, meaning that they would have been alive in intervention but dead in control.  

Another intuition for why the Fr\'{e}chet bounds are not sharp in finite samples is that they do not account for estimation error.  By holding the marginal distributions of the potential outcomes constant, they do not allow for the possibility that the marginal distributions are measured with error.  The basis of Fisher's exact test and of asymptotic inference is that even if an intervention is truly efficacious, estimates can vary and even have the wrong sign.

The maximin utility decision criterion has been criticized for being too conservative, and we show that a conservative rule based on the Fr\'{e}chet bounds was not conservative enough, thereby exposing the maximin criterion to further criticism.  As alternatives to the maximin utility decision criterion, we also consider the performance of these finite sample likelihood-based rules under decision criteria based on maximin normalized utility (i.e.minimax regret for utility instead of loss) and expected utility, as well as under a novel maximum likelihood criterion.

\begin{figure}[!hbt]
	\caption{Performance of Decision Rules under the Maximin Normalized Utility Criterion (i.e.Minimax Regret)}
	\centering
	\includegraphics[width=1\linewidth]{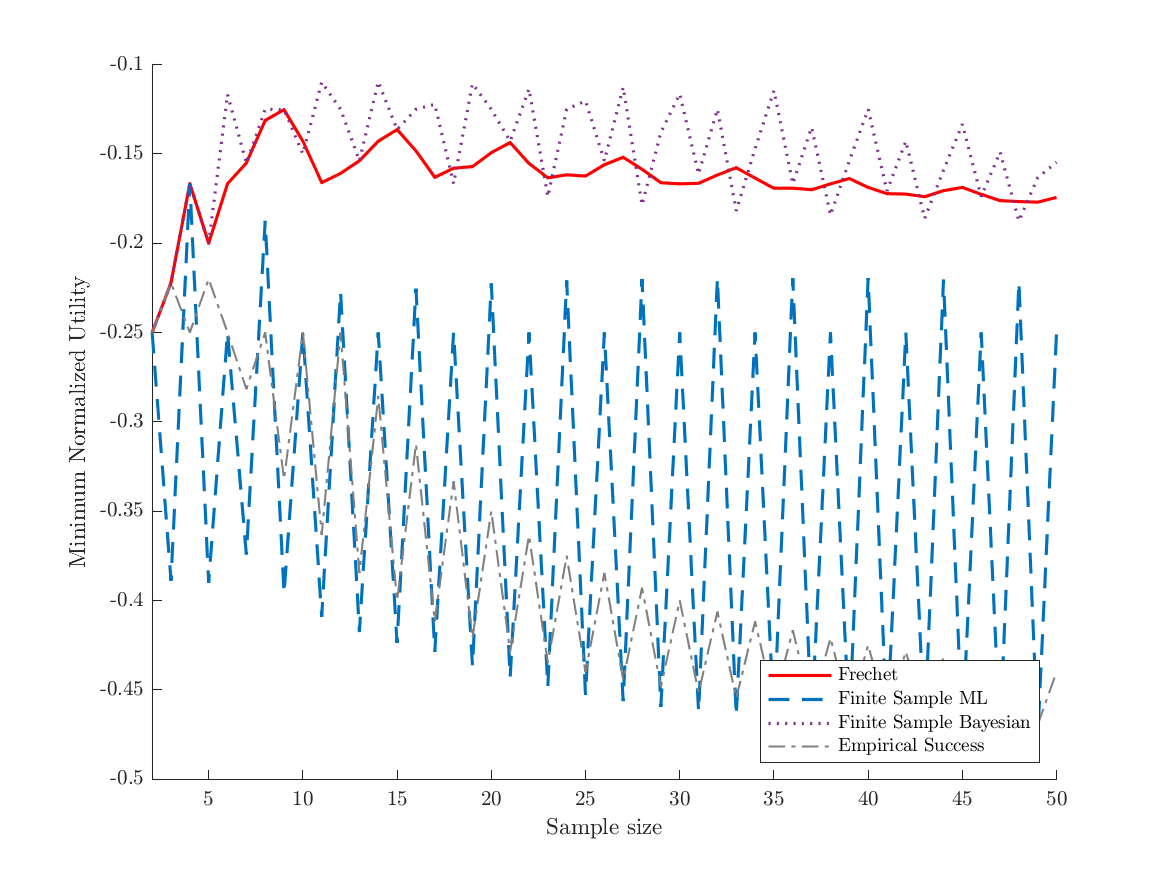}
	\label{fig:minUReg} 
\end{figure}

\begin{figure}[!hbt]
	\caption{Performance of Decision Rules under the Expected Utility Criterion}
	\centering
	\includegraphics[width=1\linewidth]{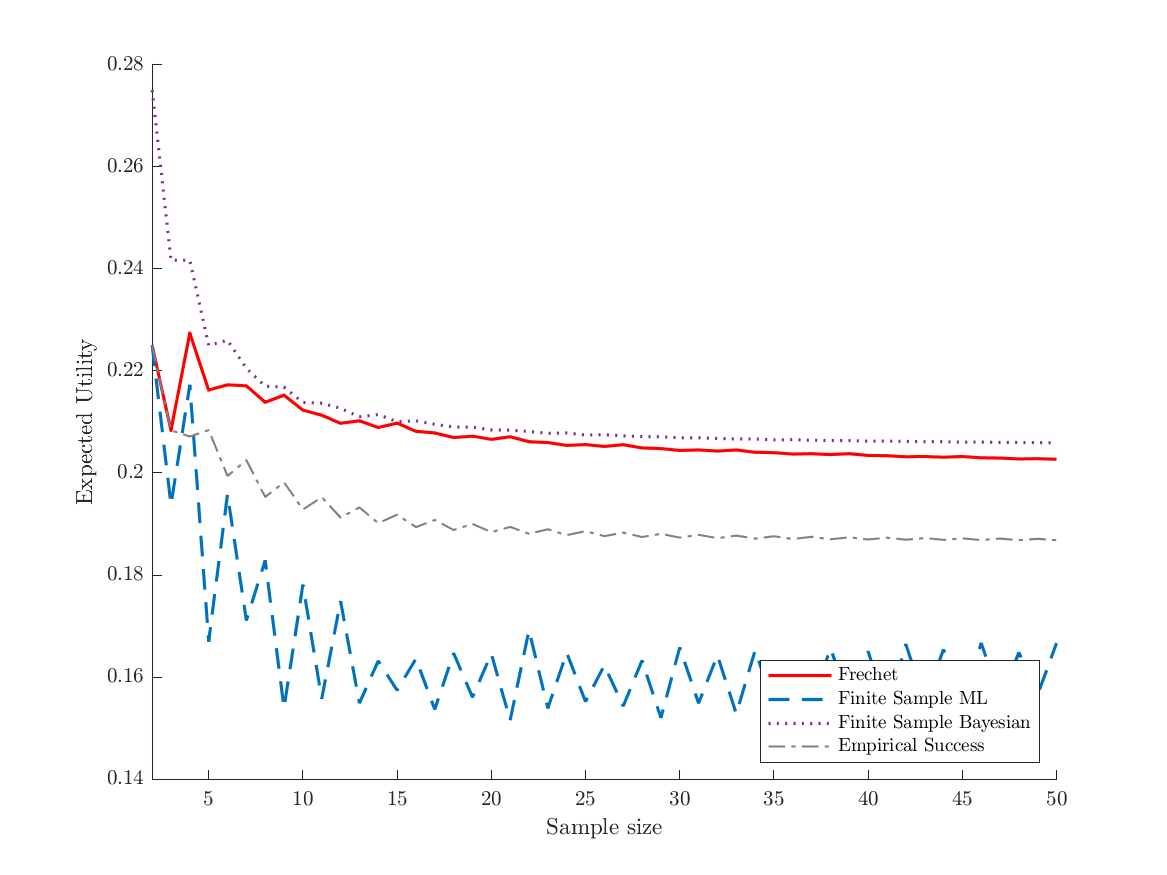}
	\label{fig:expU} 
\end{figure}

Figure \ref{fig:minUReg} shows the minimum normalized utility attained by the four decision rules in samples ranging from 2 to 50.  Our Bayesian rule shows better performance than the Fr\'{e}chet rule in terms of minimum normalized utility for many (thought not all) sample sizes.  Figure \ref{fig:expU} repeats this exercise in terms of expected utility, and as expected, the Bayesian rule performs (weakly) uniformly better than the Fr\'{e}chet rule, with large deviations in expected utility at very small sample sizes.

\begin{figure}[!hbt]
	\caption{Performance of Decision Rules under the Maximum Likelihood Criterion}
	\centering
	\includegraphics[width=1\linewidth]{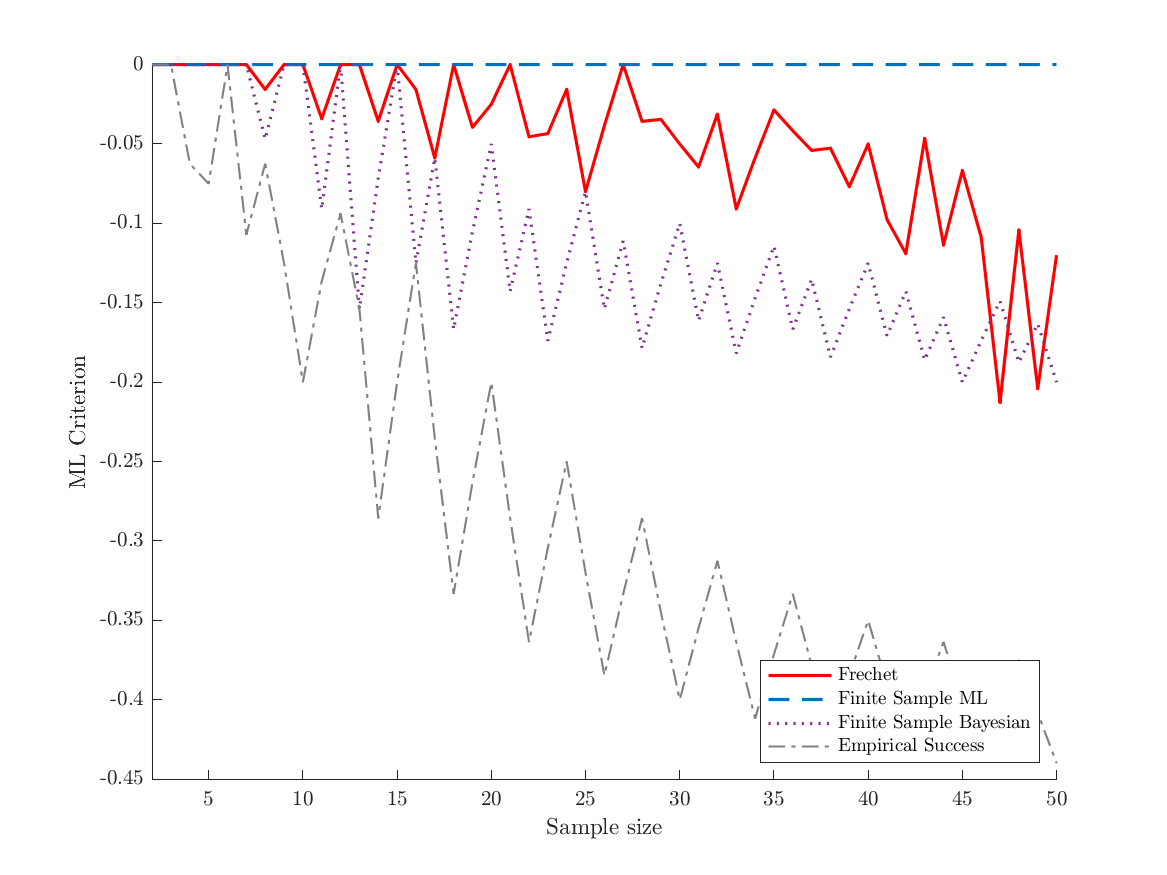}
	\label{fig:MLCrit} 
\end{figure}

Figure \ref{fig:MLCrit} shows the performance of each rule under the proposed maximum likelihood criterion.  The maximum likelihood achieves the best performance by construction.  The Fr\'{e}chet rule generically shows the best performance among the alternatives to the maximum likelihood rule that we consider.

\section{Application to a Clinical Trial of Sepsis Treatment} \label{sec:applicaiton}

Consider an actual clinical trial of 28 people that examined the impact of high dose Vitamin C on patients with sepsis \citep{zabet2016}.  In control, 9 of 14 people ($\approx 64\%$) died within 28 days, as compared with only 2 of 14 people ($\approx 14\%$) in intervention.  The Fisher exact test rejects the null hypothesis that the intervention was neither efficacious nor safe for anyone at the 2.6\% level.  Linear regression of the intervention on survival estimates an effect size of 50 percentage points and rejects the hypothesis of no effect at the 0.5\% level.

We first use the exact finite sample likelihood to answer ``Why?''  There are 4,495 ways to divide 28 people into the 4 possible types of those who would die regardless, those for whom the intervention is efficacious, those for whom the intervention is unsafe, and those who would live regardless.  The likelihood function is sparse, in the sense that only 1,260 of these possible distributions occur with positive likelihood.  Among these possible distributions, the likelihood retains some curvature.  In this case, the likelihood is uniquely maximized at a distribution with only two types of participants.  It indicates that the intervention was efficacious for 21 people, and it was unsafe for the remaining 7.  The value of the likelihood function for this maximizer is 0.153 after rounding, indicating that if an intervention is efficacious for 21 people and unsafe for 7, we will observe 9 people dead in intervention and 2 people dead in control exactly 15.3\% of the time, subject only to rounding.  Therefore, our ``best guess'' for why we observe what we did is that the intervention was efficacious for 21 people but unsafe for 7.  

Using the maximizer of the finite sample likelihood, we can also deduce how many people of each type were assigned intervention and control.  Since the maximizer rules out that any of the 12 people who lived in intervention would have lived regardless, it must have been effective for all 12 of them.  By similar logic for the people who died in control and the people who lived in died in intervention, our ``best guess'' for why we observe what we did is that it just so happened via the randomization process that more of the people for whom the intervention was effective were assigned intervention (12 vs. 9), and fewer of the people for whom the intervention was unsafe were assigned control (2 vs. 5).  Using terminology from \citet{pearl1999}, our best guess is that the intervention was ``necessary'' for the deaths of the 2 people who died in intervention because it was unsafe for them, and they would have lived without it.  Similarly, the intervention would have been ``sufficient'' for the deaths of the 5 people who lived in intervention because it was unsafe for them, so they would have died with it.

Though the maximizer of the utility function is our ``best guess,'' there are several other possibilities for ``why'' we observed what we did, and some are more likely than others.  For example, suppose that the people in the experiment were of three types instead of two.  If no one would have lived regardless and equal numbers of each type were randomized in and out, the intervention would have been unsafe for 4 people, effective for 18, and 6 would have lived regardless, with a rounded likelihood of 0.145, meaning that if these types were true, we would have observed the outcomes we did exactly 14.5\% of the time, subject only to rounding.  It's also possible that the intervention was unsafe for no one, implying that if equal numbers of each type were randomized in and out, it was effective for 14 people, 4 would have lived regardless, and 10 would have died regardless.  This possibility, which is consistent with a ``monotone response'' assumption \citep{manski1997monotone} or a LATE monotonicity assumption \citep{imbens1994} in the finite sample, is less likely, with a rounded likelihood of 0.129.  

We could also answer ``why?'' using our Bayesian decision rule.  We could use a uniform prior over the 4,495 ways that we could divide 28 people into the four types and update it according to the exact finite sample likelihood.  Doing so yields a posterior under which the intervention is expected to have been effective for 14.97 people, but it is expected to have been unsafe for 2.51 people; an expected 7.69 people would have lived regardless, and an expected 2.83 would have died regardless.

We next use statistical decision theory with the exact finite sample likelihood to answer ``What should we have done?''  By the analysis of Figure \ref{fig:minU}, a decision maker who values safety over efficacy in the proportions specified by the utility function in (\ref{eq:util}) should prefer the maximum likelihood rule in a sample of 28 by the maximin criterion.  If we use the maximum likelihood criterion, because we prefer to make our decision based on our ``best guess'' of what would happen, we should also prefer the maximum likelihood rule.  If we prefer to remain agnostic and consider the output of all rules, it is encouraging that the empirical success rule, the Fr\'{e}chet rule, our proposed finite sample Bayesian rule, and our proposed finite sample maximum likelihood rule all indicate that we should have given the intervention to everyone.  The finite sample performance of these rules must differ because the rules yield different actions for other data that we could have observed, but in this case, they all yield the same decision.

\section{Implications} \label{sec:implications}

Many new medical interventions start small.  ``First in humans'' studies of new drugs often run experiments with very small numbers of people because of safety concerns.  Phase 0, 1, and 2 trials often prioritize safety over efficacy, so much so that sometimes investigations of efficacy do not begin until Phase 3.  We provide an approach to learn about safety and efficacy simultaneously in small samples using the exact finite sample likelihood, which arises from the randomization process and is under the control of the experimenter.

We start small by proposing to use the finite sample likelihood to make decisions that prioritize safety over efficacy in small samples.  However, it is feasible to use the finite sample likelihood in large samples, especially given advances in computing power since the time of Fischer \citep{robert1996}.  The main computational challenge in large samples is not in applying the proposed rules, but rather in evaluating their performance.  The finite sample maximum likelihood rule is optimal according to the maximum likelihood criterion, and the finite sample Bayesian rule is optimal according to an expected utility criterion with a uniform prior.  Evaluating the finite sample performance of our rules in larger samples according to other criteria is an area for future work.  

A related area for future work is extending what we can learn in finite samples to other finite samples or to a population.  If an intervention is unsafe for someone in a finite sample, it must be unsafe for at least one person in the population from which the finite sample was drawn.  Extrapolation can allow us to move beyond the question of ``what should we have done in our finite sample?'' to ``what should we do in the larger population?''

Another area for future work is to think more about the implications of the maximum likelihood criterion.  Some general lessons could emerge.   As discussed by Andrew Gelman and Keith O-Rourke,  ``Awareness of commonness can lead to an increase in evidence regarding the target; disregarding commonness wastes evidence; and mistaken acceptance of commonness destroys otherwise available
evidence. It is the tension between these last two processes that drives many of the theoretical and practical controversies within statistics''  \citep{gelman2017}. In the examples that we have considered, people can be the same or different, but it is most likely that they are as similar as possible.

\setcounter{figure}{0}\renewcommand\thefigure{\Alph{subsection}.\arabic{figure}}
\setcounter{table}{0} \renewcommand\thetable{\Alph{subsection}.\arabic{table}}

\renewcommand\thesubsection{Appendix \Alph{subsection}}
\renewcommand\thesubsubsection{\thesubsection .\arabic{subsubsection}}

\titleformat{\section}
  {\bfseries\Large}
  {\thesection}{1em}{}

\titleformat{\subsection}
  {\bfseries\large}
  {\thesubsection}{1em}{\normalfont\large}

\titleformat{\subsubsection}
  {\bfseries}
  {\thesubsubsection}{1em}{\normalfont}

\vspace{5mm}
\section*{Appendix} \label{sec:appendix}

\addcontentsline{toc}{section}{Appendix}

\subsection{Likelihood Derivation} \label{sec:app_likelihood}

Let 
$\Ib \equiv \big( I_{1,1}, I_{1,0}, I_{0,1}, I_{0,0} \big)$ be a random vector whose elements represent the numbers of individuals randomized into intervention among those who would live regardless, those for whom the intervention would be efficacious, those for whom the intervention would be unsafe, and those who would die regardless. These random variables jointly follow a multivariate hypergeometric distribution, where $m$ total draws are made from four groups whose sizes are represented by $\thetab$:
\begin{align*}
\Prob \Big( I_{1,1} = i_{1,1}, I_{1,0} = i_{1,0}, 
	 I_{0,1} = i_{0,1}, I_{0,0} = i_{0,0} \mid \thetab \Big)
= \frac{ 
	\binom{\theta_{1,1}}{i_{1,1}} \binom{\theta_{1,0}}{i_{1,0}}
		\binom{\theta_{0,1}}{i_{0,1}} \binom{\theta_{0,0}}{i_{0,0}} }
	{ \binom{n}{m} }
\end{align*}

The observable data $\Xb$ can be expressed in terms of the latent $\Ib$ variables and the distribution of potential outcomes $\thetab$ by observing that each individual randomized into the intervention group with outcome $Y=1$ must have either been someone who would have lived regardless or for whom the intervention was efficacious: 
$X_{I1} = I_{1,1} + I_{1,0}$.  Then, since only $m$ individuals are randomized into intervention, we must have that 
$X_{I0} = m - \big( I_{1,1} + I_{1,0} \big)$.  In the control group, those observed with outcome $Y=1$ must have either been someone who would have lived regardless or for whom the intervention would have been unsafe:
$X_{C1} = \big( \theta_{1,1} - I_{1,1} \big) 
	+ \big( \theta_{0,1} - I_{0,1} \big)$.  And, since only $m-n$ individuals are randomized into control, we must have that
$X_{C0} = (n - m) - \big( \theta_{1,1} - I_{1,1} \big)
	- \big( \theta_{0,1} - I_{0,1} \big)$.  
Thus, we can write the probability of the observed data $\Xb$ conditional on the distribution of potential outcomes $\thetab$ as:
\begin{align*}
\Prob \big( \Xb = \boldsymbol{x} \mid \thetab \big)
&= \Prob \Big( 
	I_{1,1} + I_{1,0} = x_{I1},\ 
	\big( \theta_{1,1} - I_{1,1} \big) 
		+ \big( \theta_{0,1} - I_{0,1} \big) = x_{C1}
	\mid \thetab \Big) \\
&= \Prob \Big( 
	I_{1,1} + I_{1,0} = x_{I1},\ 
	I_{1,1} + I_{0,1} = \theta_{1,1} + \theta_{0,1} - x_{C1}
	\mid \thetab \Big)
\end{align*}

Each realization of $\Xb$ can be produced from multiple realizations of $\Ib$.  Thus, to find the probability of a realization of $\Xb$, we sum together the probabilities of each 
realization of $\Ib$ that could have produced it.  We can index these realizations through the realization $i$ of $I_{1,1}$ ranging from $0$ to $\theta_{1,1}$ and solving the following system of equations for the elements of $\Ib$:
\begin{align*}
I_{1,1} + I_{1,0} &= x_{I1}, \\
I_{1,1} + I_{0,1} &= \theta_{1,1} + \theta_{0,1} - x_{C1}, \\
I_{1,1} + I_{1,0} + I_{0,1} + I_{0,0} &= m, \\
I_{1,1} &= i
\end{align*}
Rearranging yields
\begin{align*}
I_{1,1} &= i \\
I_{1,0} &= x_{I1} - i \\
I_{0,1} &= \theta_{1,1} + \theta_{0,1} - x_{C1} - i \\
I_{0,0} &= m + x_{C1} + i - \theta_{1,1} - \theta_{0,1} - x_{I1}  
\end{align*}
The probability of a realization of $\Xb$ is just the sum of the probability of each of these realizations of $\Ib$:
\begin{align*}
\Prob \big( \Xb = \boldsymbol{x} \mid \thetab \big)
&= \sum_{i=0}^{\theta_{1,1}} \Prob \Big(
	I_{1,1} = i, \\
	& \qquad\qquad I_{1,0} = x_{I1} - i, \\
	& \qquad\qquad I_{0,1} = \theta_{1,1} + \theta_{0,1} 
		- x_{C1} - i, \\
	& \qquad\qquad I_{0,0} = m + x_{C1} + i - \theta_{1,1} 
		- \theta_{0,1} - x_{I1}  
	\mid \thetab \Big).
\end{align*}
Substituting the distribution of $\Ib$ from above yields our likelihood expression in (\ref{eq:likelihood}):
\begin{align*}
\Prob \big( \Xb = \boldsymbol{x} \mid \thetab \big)
&= \sum_{i=1}^{\theta_{1,1}}
	\binom{ \theta_{1,1} }{ i } \\
	&\qquad \qquad \times 
		\binom{ \theta_{1,0} }{ x_{I1} - i }  \\
	&\qquad \qquad \times 
		\binom{ \theta_{0,1} }
			{ \theta_{1,1} + \theta_{0,1} - x_{C1} - i }  \\
	&\qquad \qquad \times
		\binom{ \theta_{0,0} }
			{ m + x_{C1} + i - \theta_{1,1} 
				- \theta_{0,1} - x_{I1} }
	\bigg/ \binom{n}{m}
\end{align*}

\subsection{Fr\'{e}chet Rule as Estimated Average Intervention Effect Cutoff Rule} \label{sec:app_cutoff}

Under the utility specification in (\ref{eq:util}), the Fr\'{e}chet rule prescribes action if
\begin{align*}
\min_{ p \in \widehat{FH}(\Xb) } 
	\Big( \frac{1}{2} p_{1,0} - p_{0,1} \Big)
> \min_{ p \in \widehat{FH}(\Xb) }
	\Big( p_{0, 1} - \frac{1}{2} p_{1,0} \Big)
\end{align*}
Applying the definition of the set $\widehat{FH}(\Xb)$ in (\ref{eq:FH}), the left side of the above inequality can be expressed as a maximization over $p_{0,1}$:
\begin{align*}
\min_{ p \in \widehat{FH}(\Xb) }
	\bigg( \frac{1}{2} p_{1,0} - p_{0,1} \bigg)
&= \min_{ p_{0,1} \in [ \underline{p}, \bar{p} ] }
	\bigg( \frac{1}{2} \big( \hat{p}_I(\Xb) - \hat{p}_C(\Xb) + p_{0,1} \big)
		- p_{0,1} \bigg) \\
&= \min_{ p_{0,1} \in [ \underline{p}, \bar{p} ] }
	\bigg( \frac{1}{2} \Big( 
		\hat{p}_I(\Xb) - \hat{p}_C(\Xb) \big)
		- p_{0,1} \Big) \bigg),
\end{align*}
where 
$\underline{p} \equiv \max \big\{ 
	- \big( \hat{p}_I(\Xb) - \hat{p}_C(\Xb) \big), 0\big\}$
and
$\bar{p} \equiv \min \big\{ \hat{p}_C(\Xb), 1 - \hat{p}_I(\Xb) \big\}$.
This expression is decreasing in $p_{0,1}$, so it takes its minimum value at $p_{0,1} = \bar{p}$: 
\begin{align*}
\min_{ p \in \widehat{FH}(\Xb) }
	\Big( \frac{1}{2} p_{1,0} - p_{0,1} \Big)
&= \frac{1}{2} \Big( \hat{p}_I - \hat{p}_C 
	- \min \big( \hat{p}_C(\Xb), 1 
		- \hat{p}_I(\Xb) \big) \Big).
\end{align*}

The right side of the inequality can be transformed similarly:
\begin{align*}
\min_{ p \in \widehat{FH}(\Xb) }
	\bigg( p_{0,1} - \frac{1}{2} p_{1,0} \bigg)
&= \min_{ p_{0,1} \in [ \underline{p}, \bar{p} ] }
	\bigg( p_{0,1} - \frac{1}{2} \big( \hat{p}_I(\Xb) - \hat{p}_C(\Xb) 
		+ p_{0,1} \big) \bigg)  \\
&= \min_{ p_{0,1} \in [ \underline{p}, \bar{p} ] }
	\bigg( \frac{1}{2} \Big( p_{0,1} 
		- \big( \hat{p}_I(\Xb) - \hat{p}_C(\Xb) \big) \Big) \bigg) 
\end{align*}
This expression is increasing in $p_{0,1}$, so it takes its minimum at $p_{0,1} = \underline{p}$:
\begin{align*}
\min_{ p \in \widehat{FH}(\Xb) }
	\bigg( p_{0,1} - \frac{1}{2} p_{1,0} \bigg)
&= \max \Big( - \big( \hat{p}_I(\Xb) 
	- \hat{p}_C(\Xb) \big), 0 \Big)
- \big( \hat{p}_I(\Xb) - \hat{p}_I(\Xb) \big)
\end{align*}

When the estimated average intervention effect 
$ \hat{p}_I(\Xb) - \hat{p}_C(\Xb) $
is positive, the inequality simplifies to
\begin{align*}
\hat{p}_I(\Xb) - \hat{p}_C(\Xb) 
	> \frac{1}{2} \min \big( \hat{p}_C(\Xb), 
		1 - \hat{p}_I(\Xb) \big)
\end{align*}
When the estimated average intervention effect is negative, the inequality simplifies to
\begin{align*}
\hat{p}_I(\Xb) - \hat{p}_C(\Xb) 
	> \min \big( \hat{p}_C(\Xb), 
		1 - \hat{p}_I(\Xb) \big),
\end{align*}
which is never satisfied for 
$ \hat{p}_I(\Xb) - \hat{p}_C(\Xb) < 0 $.
So, we conclude that the Fr\'{e}chet rule prescribes action whenever the estimated average intervention effect exceeds the positive threshold value 
$ \frac{1}{2} \min \big( \hat{p}_C(\Xb), 
	1 - \hat{p}_I(\Xb) \big)$.

\newpage
\singlespacing
\bibliographystyle{chicago}
\bibliography{decide}

\end{document}